\documentclass[twocolumn]{jpsj2} 
\def\mbf#1{\text{\boldmath $#1$}}
\def\e{\rm{e}}
\def\i{\text{i}}
\def\ket#1{|#1\rangle }
\def\bra#1{\langle #1|}
\def\figsize{0.6\hsize}

\def\paragraph#1{\section{#1}}
\begin{document}
\title{Topological Aspects of Surface States in Semiconductors}
\author{Yoshihiro \textsc{Kuge}$^{1}$\thanks{E-mail address: kuge@pothos.t.u-tokyo.ac.jp}, Isao \textsc{Maruyama}$^{1}$\thanks{E-mail address: maru@pothos.t.u-tokyo.ac.jp} and Yasuhiro \textsc{Hatsugai}$^{1,2}$\thanks{E-mail address: hatsugai@sakura.cc.tsukuba.ac.jp}}
\inst{$^{1}$Department of Applied Physics, Univ. of Tokyo, 7-3-1 Hongo, Bunkyo-ku Tokyo 113-8656, JAPAN \\
$^{2}$Institute of Physics, Univ. of Tsukuba, 1-1-1 Tennodai, Tukuba Ibaraki 305-8571, JAPAN
}
\abst{
 Topological aspects of
 surface states in semiconductors
 are studied by an adiabatic deformation which connects
 a realistic system and a decoupled
 covalent-bond model.
 Two topological invariants are focused.
 One is a quantized Berry phase,
 and the other is a number of the edge states.
 A winding number as another topological invariant is
 also considered.
 The surface states of Si and Ge at
 (111), (110), and (100) surfaces
 are classified by the
 topological invariants.
 Surface states of the GaAs as heterosemiconductors are also discussed.
}
\kword{semiconductor surface, Berry phase, edge state, dangling bond, Ge, Si, GaAs}
\date{\today}
\maketitle
 \paragraph{Introduction}
 For a few decades the topological feature which does not depend on
 the detail of systems
 has been attracting attention in the condensed matter physics.
 One of most remarkable examples is a quantum Hall effect (QHE)\cite{Klitzing_1980},
 where a topological invariant known as the Chern number
 \cite{Thouless_Kohmoto_Nightingale_den_Nijs_1982}
 has been observed experimentally.
 The Chern numbers are defined for a bulk state without boundaries.
 In the QHE, the edge states which are characteristic
 for a system with boundary also have a significant topological
 importance~\cite{Hatsugai_1993}.

 Edge states or surface states,
 which localize at a boundary of a system,
 have significant effects in physical phenomena;
 e.g., 
 the zero bias conductance peak structure in anisotropic
 superconductivity
 \cite{Tanaka_Kashiwaya_1995,Kashiwaya_Tanaka_1995},
 single-layer graphite
 \cite{Fujita_1996},
 and the boundary local moment at an edge of hexagonally bonded honeycomb sheets consisting of B, N, and C atoms
 \cite{Okada_Oshiyama_2001}.
 Especially, in a graphene, it is now widely understood that there exist
 important edge states which are related to the chiral symmetry of the
 system and the Dirac cone spectrum.
 In metal-semiconductor interfaces,
 an interface state, known as
 the metal induced gap state,
 has an important role in understanding
 Fermi-level pinning~\cite{Heine_1965}.

 Moreover, in realistic materials,
 there exists
 a surface reconstruction or a surface relaxation,
 which is  sensitive to the detail of systems.
 However, we shall discuss that
 topological invariants which is independent of 
 the detail of models
 is useful for understanding the surface states of 3 dimensional semiconductors.
 A candidate of topological invariants for semiconductors
 is the quantized Berry phase which has been
 proposed as a local order parameter for clarification of the phase of the gapped quantum liquids.
 It has been demonstrated that quantization of this local order parameter has an advantage over the usual order parameter
 in a frustrated spin system\cite{Hatsugai_2006},
 and a strongly correlated system\cite{Maruyama_Hatsugai_2007}.

 In this paper, 
 the quantized Berry phase is
 evaluated for
 surface states in typical semiconductors
 such as Si, Ge, and GaAs.
 We note that the Berry (or Zak) phase also appears
 in the King-Smith-Vanderbilt formula
 in the theory of macroscopic polarization.\cite{King-Smith_Vanderbilt_1993,Resta_2000}
 In our case the Berry phase is defined for a 1-dimensional system with
 2-dimensional wave vectors to characterize the surface, and
 quantized due to a anti-unitary symmetry.
 In addition,
 we focused on the number of the edge states and a winding number
 as the other topological invariants, following Ref.\citen{Ryu_Hatsugai_2002}.
 For simplification,
 we limit ourselves to surface states in ideal surfaces understood as
 Shockley states\cite{Shockley_1939}.
 Shockley states at a metal-semiconductor contact
 are closely-linked to dangling bonds.\cite{Bardeen_1947}
 Although we does not consider a surface reconstruction,
 the study on surface states at ideal surfaces sheds a light on real materials
 in the sense that
 the number of dangling bonds in a respective surface
 acts as a trigger for surface-dependent physics;
 for  example,
 $(7 \times 7)$ structures on (111) surface of Si\cite{Takayanagi_1985}
 and Ge\cite{Becker_Golovchenko_Swartzentruber_1985},
 $(16 \times 2)$ structures on (110) surface of Si\cite{Yamamoto_Ino_Ichikawa_1986}
 and Ge\cite{Noro_Ichikawa_1985},
 and $c(4\times 2)$ structures on (100) surface of 
 Si\cite{Inoue_1994} 
 and Ge\cite{Yoshimoto_2000}. 
 \paragraph{Model and Results}
 To consider
 (111), (110), and (100) surfaces of Si, Ge, and GaAs
 as typical IV and III-V semiconductors,
 we use tight binding models determined by the Slater and Koster's
 energy integrals\cite{Slater_Koster_1954}.
 We also consider a spin-orbit coupling that is important for Ge.
 We use the tight binding and spin-orbit coupling parameters of
 Grosso and Piermarocchi\cite{Grosso_Piermarocchi_1995}
 for Ge,
 those of
 Klimeck {\it et. al.} \cite{Klimeck_2000}
 for Si,
 and those of Boykin {\it et. al.}\cite{Boykin_2002} for GaAs.
 Note that a $sp^{3}$ model is considered for Ge, 
 and a $sp^{3}s^{\ast}$ model for Si, and GaAs.

\begin{figure}[hbtp]
  \begin{center}
   \includegraphics[width=\figsize,angle=0]{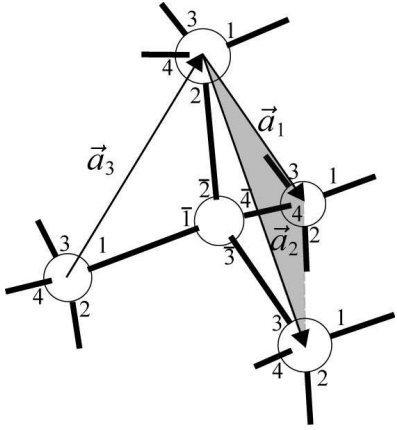}
  \end{center}
  \caption{Model of a diamond structure. Spheres stand for atoms, and lines
  stand for $sp^{3}$ hybridized orbitals. A shade stands for (111)
  surface, and $\vec{a}_{n}$ stands for basis lattice vectors relative to (111) surface.}
  \label{fig.model}
 \end{figure}

 To define a surface-dependent Berry phase,
 let us start from the definition of a bulk Hamiltonian.
 A position of a center of a unit cell is given by
 $\mbf{r}_{i} = \sum_{n = 1}^{3} i_{n} \mbf{a}_{n},\;$
 $i_n \in \mathbb{Z}$,
 where $\mbf{a}_{1}$, $\mbf{a}_{2}$, and
 $\mbf{a}_{3}$ are primitive
 vectors.
 Primitive reciprocal lattice vectors $\mbf{b}_{n}$
 are defined by
 $\mbf{a}_{n}\cdot\mbf{b}_{m}=2\pi\delta_{nm}$,
 where $\delta_{nm}$ is the Kronecker delta.
 A wavenumber vector $\mbf{k}$ is given by
 $\mbf{k} = \sum_{n=1}^{3}k_{n}\mbf{b}_{n}$.
 The bulk Hamiltonian in three dimensional
 periodic boundary condition (PBC)
 $\mathcal{H}_{\text{PBC}}$ is given by
 \begin{equation}
  \mathcal{H}_{\text{PBC}}=
   \sum_{\mbf{k}}
   \sum_{\beta,\beta'}
   c_{\mbf{k},\beta}^{\dagger}
   \left[
    H_{\text{PBC}}(\mbf{k})
   \right]_{\beta,\beta'}
   c_{\mbf{k},\beta'}.
 \end{equation}
 Here, the label $\beta=(\alpha,\sigma)$ distinguishes spin-dependent orbitals in a unit cell.
 We suppose two atoms in a unit cell
 and denote orbitals as $\alpha=1,\bar{1},2,\bar{2},\ldots, N,\bar{N}$,
 where $N$ is a number of orbitals in a atom
 and $\bar{\alpha}$ on a atom means an orbital opposite to an orbital $\alpha$ on the other atom
 (See Fig.~\ref{fig.model}).
 Then, $H_{\text{PBC}}(\mbf{k})$ is
 a $4N\times4N$ matrix.

 Next, we suppose that $\mbf{a}_{1}$ and $\mbf{a}_{2}$ span a plane
 parallel to a surface of a semiconductor as a shaded plane shown in Fig.~\ref{fig.model}.
 2-dimensional wavenumber vector is given by
 $\mbf{k}_{\|}=\sum_{n=1}^{2}k_{n}\mbf{b}_{n}$.
 Then, the bulk Hamiltonian is written as
 \begin{equation}
  \mathcal{H}_{\text{PBC}}=
   \sum_{\mbf{k}_{\|}}
   \sum_{n,n'}
   c_{\mbf{k}_{\|},n}^{\dagger}
   \left[
    H_{\text{PBC}}(\mbf{k}_{\|})
   \right]_{n,n'}
   c_{\mbf{k}_{\|},n'},
 \end{equation}
 where the label $n$ is defined as
 $n=(i_{3},\beta)$ and
 $H_{\text{PBC}}(\mbf{k}_{\|})$ is
 a $4NL\times4NL$ matrix, and
 $i_{3}\in[1,L]$.
 $L$ is a number of layers.
 To consider the surface,
 we introduce the open boundary condition (OBC) for the one dimensional system in the $\mbf{a}_{3}$ direction.
 The matrix ${H}_{\text{OBC}}(\mbf{k}_{\|})$ is given by truncating
 ${H}_{\text{PBC}}(\mbf{k}_{\|})$.
 For example, a natural way to truncation is to prohibit all the matrix
 elements across $L$.
 We define a boundary-hopping matrix
 $H_{\text{B}}(\mbf{k}_{\|})=H_{\text{PBC}}(\mbf{k}_{\|})-H_{\text{OBC}}(\mbf{k}_{\|})$
 as
 \begin{equation}
  \left[H_{\text{B}}(\mbf{k}_{\|})\right]_{n,n'}=
   \left\{
    \begin{array}{ll}
     \left[H_{\text{PBC}}(\mbf{k}_{\|})\right]_{n,n'}&[i_{3},i_{3}']\ni L\\
     0&\text{others}\\
    \end{array}
   \right..
 \end{equation}
 ${H}_{\text{B}}(\mbf{k}_{\|})$ is $4NL\times4NL$ matrix.
 This matrix represents all hopping elements across the boundary
 of the system.

 To define the Berry phase,
 we introduce a twist angle
 $\theta$ in a hopping term across the boundary as
 $\displaystyle c_{\mbf{k}_{\|},n}^\dagger c_{\mbf{k}_{\|},n'}\rightarrow\e^{\i\theta}c_{\mbf{k}_{\|},n}^\dagger c_{\mbf{k}_{\|},n'}$.
 In detail,
 the twist angle $\theta$ is introduced in a selected element $nn'$ of ${H}_{\text{B}}(\mbf{k}_{\|})$
 and we denote it as ${H}_{\text{B}}(\theta, \mbf{k}_{\|})$.
 Then, the Berry phase can be defined as
 \begin{equation}
   \gamma_{nn'}(\mbf{k}_{\|}) = \int_0^{2\pi} \bra{gs}\partial_{\theta}\ket{gs}d\theta
   ,
 \end{equation}
 where  $|gs(\theta) \rangle$ is the half-filled ground state of ${H}_{\text{PBC}}(\theta, \mbf{k}_{\|})
 = {H}_{\text{OBC}}(\mbf{k}_{\|}) + {H}_{\text{B}}(\theta, \mbf{k}_{\|})$.
 Especially,
 we denote main elements of the Berry phase as $\gamma_{\alpha\sigma} = \gamma_{(L\alpha\sigma)(1\bar{\alpha}\sigma)}$.
 The Berry phase is quantized because
 the one-dimensional system at $\mbf{k}_{\|}$ defined by ${H}_{\text{PBC}}(\mbf{k}_{\|})$
 has a inversion symmetry at the boundary:
 $(i_3, \alpha) \leftrightarrow (L+1-i_3, \bar{\alpha})$.
 When we denote the inversion as $U$
 and the complex conjugate as $K$,
 the anti-unitary operator $\Theta = K U$ is commutable with the Hamiltonian.
 Then, the Berry phase turns out to be quantized in the same way as in Ref.\cite{Hatsugai_2006}.

\begin{figure}[hbtp]
   \begin{center}
  (a)\\
  \includegraphics[width=\figsize]{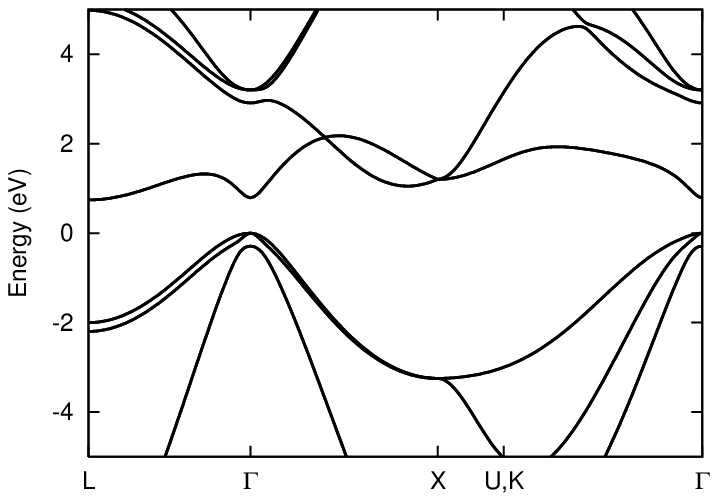}\\
  (b)\\
  \includegraphics[width=\figsize]{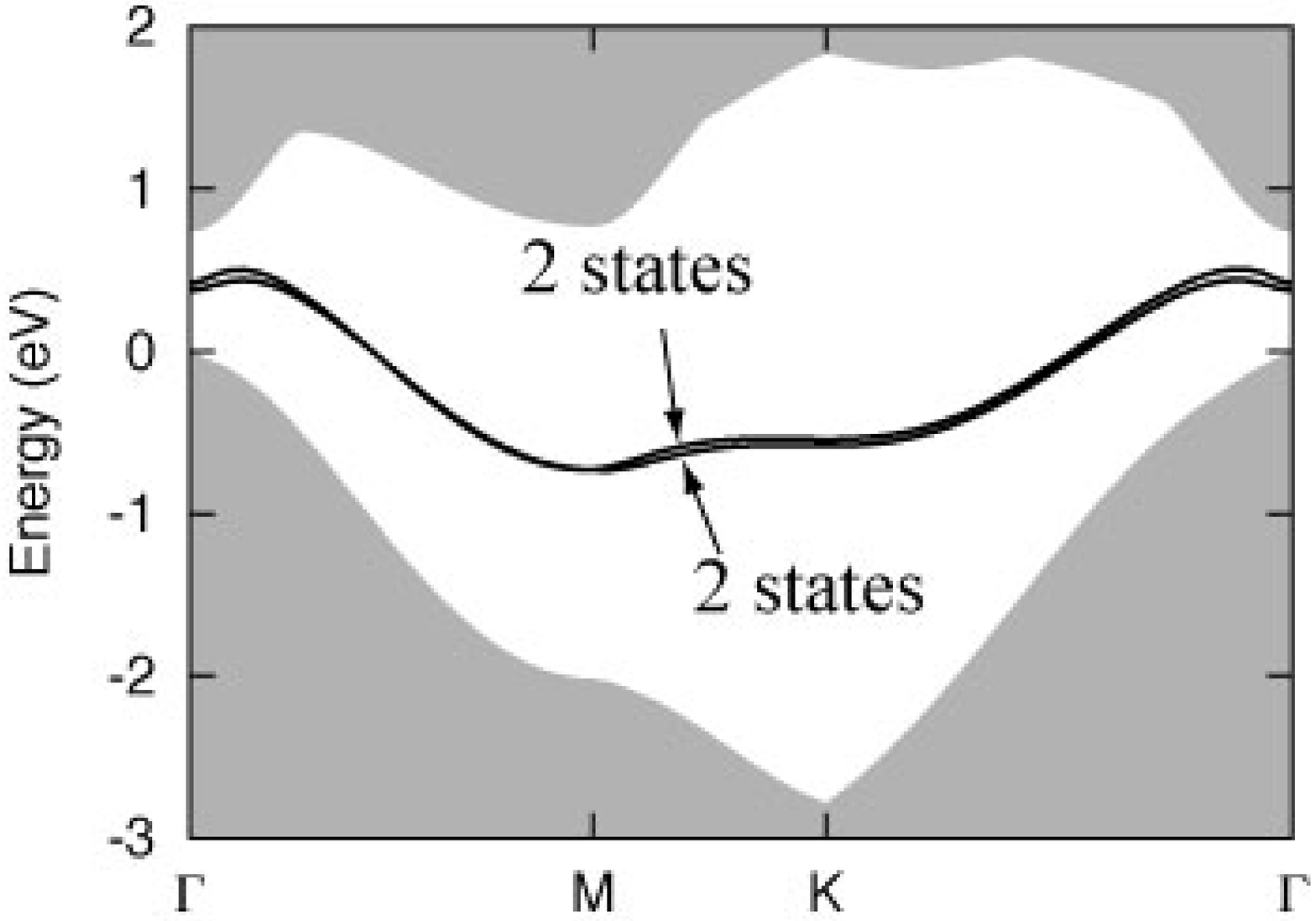}\\
  \caption{(a) A band structure of bulk Ge.
  (b) Energy spectrum of Ge (111) surface.
  The white area denotes band gap of bulk Ge and
  lines do edge states in Ge (111).
  The notations of points in relative surface Brillouin zone are
  according to Ivanov {\it et. al.}\cite{Ivanov_1980}.
  The number of the degeneracy of the edge states is four
  including spin degrees of freedom.
  }
  \label{fig.1}
\end{center}     
\end{figure}

 Figure~\ref{fig.1} (a)
 shows the band structure of $\mathcal{H}_{\text{PBC}}$
 of Ge
 and reproduce the indirect gap.\cite{Kittel_1996}
 After introducing the (111) surface,
 the band diagram of $\mathcal{H}_{\text{OBC}}$
 has
 edge states in the band gap, which are doubly-degenerated per spin,
 as shown in Fig.~\ref{fig.1} (b).
 The Berry phase of Ge (111) surface is actually quantized
 and does not depend on $\mbf{k}_{\|}$.
 It turns out to be $\gamma_{1\uparrow}=\gamma_{1\downarrow}=\pi$ on the strongest hopping in ${H}_{\text{B}}$
 and zero on the others.
 We note that the number of candidates of $\gamma_{nn'}$ is 48
 for the parameters we used.

 The quantized Berry phase and the number of edge states $n_e$
 are topological invariant in the following adiabatic transformation.
 We first modify the
 spin-orbit coupling term
 adiabatically to zero preserving the energy gap open, that is,
 without changing the topological invariants.
 At this stage the Hamiltonian is spin decoupled as $\mathcal{H}=\mathcal{H}_\uparrow\oplus \mathcal{H}_\downarrow$.
 Now we ignore the spin index below for simplification.
 Next, we can modify the hopping terms between different orbitals
 and obtain a Hamiltonian as the sum of two band Hamiltonians
 \begin{math}
  \mathcal{H}=
   \mathcal{H}^{(1)}
   \oplus\cdots\oplus
   \mathcal{H}^{(N)},
 \end{math}
 where $\mathcal{H}^{(\alpha)}$ involves hopping terms between
 $\alpha$ and $\bar{\alpha}$ only.
 For this process, it is also possible to hold the gap open.
 After the Fourier transformation of the $\mbf{a}_{3}$ direction,
 $\mathcal{H}_{\text{PBC}}^{(\alpha)}$ can be written as a 2 $\times$ 2 matrix
 $H_{\text{PBC}}^{(\alpha)} (\mbf{k}=(\mbf{k}_{\|},k_{3}))$.
 We can expand $H_{\text{PBC}}^{(\alpha)}$ by the Pauli matrices
 $\mbf{\sigma}$ by
 \begin{equation}
  H_{\text{PBC}}^{(\alpha)} (\mbf{k}) =
  \mbf{R}^{(\alpha)} (\mbf{k})
   \cdot \mbf{\sigma}
   + R_0^{(\alpha)} (\mbf{k}) {\sigma_0},
 \end{equation}
 where $\sigma_0$ is the 2 $\times$ 2 identity matrix.
 This equation indicates a one to one mapping between
 $H_{\text{PBC}}^{(\alpha)} $
 and a four dimensional vector $(R_0, \mbf{R})$.
 Therefore,
 for fixed $\mbf{k}_{\|}$,
 a loop $\mathcal{L}_{\alpha}$ in $\mathbb{R}^{4}$ space is given
 as $k_{3}$ varies over $S^{1}=[0, 1)$.
 We note that the four dimensional vectors at $k_{3}=0$ and $k_{3}=1$ are the same.
 Mathematical representation is given by
 \begin{equation}
  \mathcal{L}:k_{3}\in S^{1}\rightarrow (R_0, \mbf{R})\in\mathbb{R}^{4}.
 \end{equation}
 Thirdly, we can modify $R_0(\mbf{k}_{\|},k_3)$ to zero preserving the gap open.
 For the arbitrary surface of Si or Ge,
 $R^{(\alpha)}_{z}(\mbf{k}_{\|},k_3)$ is zero.
 The condition that $\mathcal{L}_{\alpha}$ is restricted in a two-dimensional plane
 is called as a chiral symmetry\cite{Ryu_Hatsugai_2002}.
 The loop is given by
 \begin{equation}
  \mathcal{L}:k_{3}\in S^{1}\rightarrow
   (R_{x}, R_{y})
  \in\mathbb{R}^{2}.
 \end{equation}

\begin{figure}[htbp]
   \begin{center}
  \includegraphics[width=\figsize]{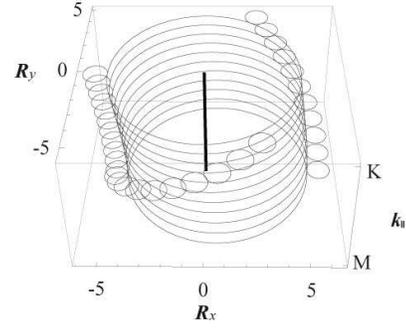}
  \caption{Two-dimensional loops $\mathcal{L}_{\alpha}(\mbf{k}_{\|})$ of the Hamiltonian with the
  chiral symmetry for the Ge (111) surface.
  The vertical axis shows $\mbf{k}_{\|}$ and the other axes show
  $xy$ elements of the loops.
  The vertical axis equals to a part of horizontal axis in
  Fig.~\ref{fig.1} (b).
  Vertical line between two solid circles denotes the origin for each $\mbf{k}_{\|}$.
  Each loops with fixed $\mbf{k}_{\|}$ locates on plane parallel to $xy$
  plane.
  Each plane has four loops related to four Hamiltonians.
  On each plane, one loop encloses the origin and the others do not.
  }
  \label{loop_Ge}
   \end{center}
 \end{figure}

 For a 2-dimensional loop, we can define a winding number
 as in Ref.~\citen{Ryu_Hatsugai_2002}.
 This is the third topological invariant we considered in the paper.
 The winding number $\mathcal{W}_{\alpha}(\mbf{k}_{\|})$ is
 defined as the total number of times that the loop travels
 counterclockwise around the origin in the $xy$ plane,
 and determines whether $H_{\text{OBC}}^{(\alpha)}(\mbf{k}_{\|})$
 has edge states or not.
 Figure~\ref{loop_Ge}
 shows $\mathcal{L}_{\alpha}(\mbf{k}_{\|})$
 of Ge (111) surface.
 In each plane vertical to the $z$ axis of Fig.~\ref{loop_Ge},
 four loops with fixed $\mbf{k}_{\|}$ are displayed
 and one of these loops encloses the origin, $R_{x}=R_{y}=0$ at each $\mbf{k}_{\|}$,
 i.e., the winding number of a link is $\mathcal{W}_{1}(\mbf{k}_{\|})=1$ for arbitrary $\mbf{k}_{\|}$,
 which clarifies existence of zero energy edge states.
 The other loops do not enclose for each $\mbf{k}_{\|}$, i.e., $\mathcal{W}_{\alpha=2,3,4}(\mbf{k}_{\|})=0$,
 which is consistent with absence of zero energy edge states for each $\mbf{k}_{\|}$.
 Note that zero energy edge states are doubly degenerated (left and right).
 This result from the winding number corresponds to the number of edge
 states $n_e=2$ per spin.

 Finally, we modify hopping parameters except for
 hopping between face-to-face orbitals which locate on nearest neighbor
 atoms to zero adiabatically.
 We call this decoupled Hamiltonian as a covalent-bond model $\mathcal{H}_{CB}$.
 Edge states of $\mathcal{H}_{CB}$ with the OBC corresponds to dangling bonds,
 where dangling bonds are defined as the covalent bonds which are cut by the surface.
 Of course,
 topological invariants $\gamma$ and $n_e$ are
 unchanged between the adiabatic deformation.
 Then, $\gamma_{nn'}=\pi$ implies that a covalent-bond bond exists on the link $nn'$
 and the integer $n_e$ corresponds to the number of dangling bonds.

 \paragraph{Summary of the other results}
 \begin{figure}[htbp]
   \begin{center}
  (a) Ge (110)\\
  \includegraphics[width=\figsize]{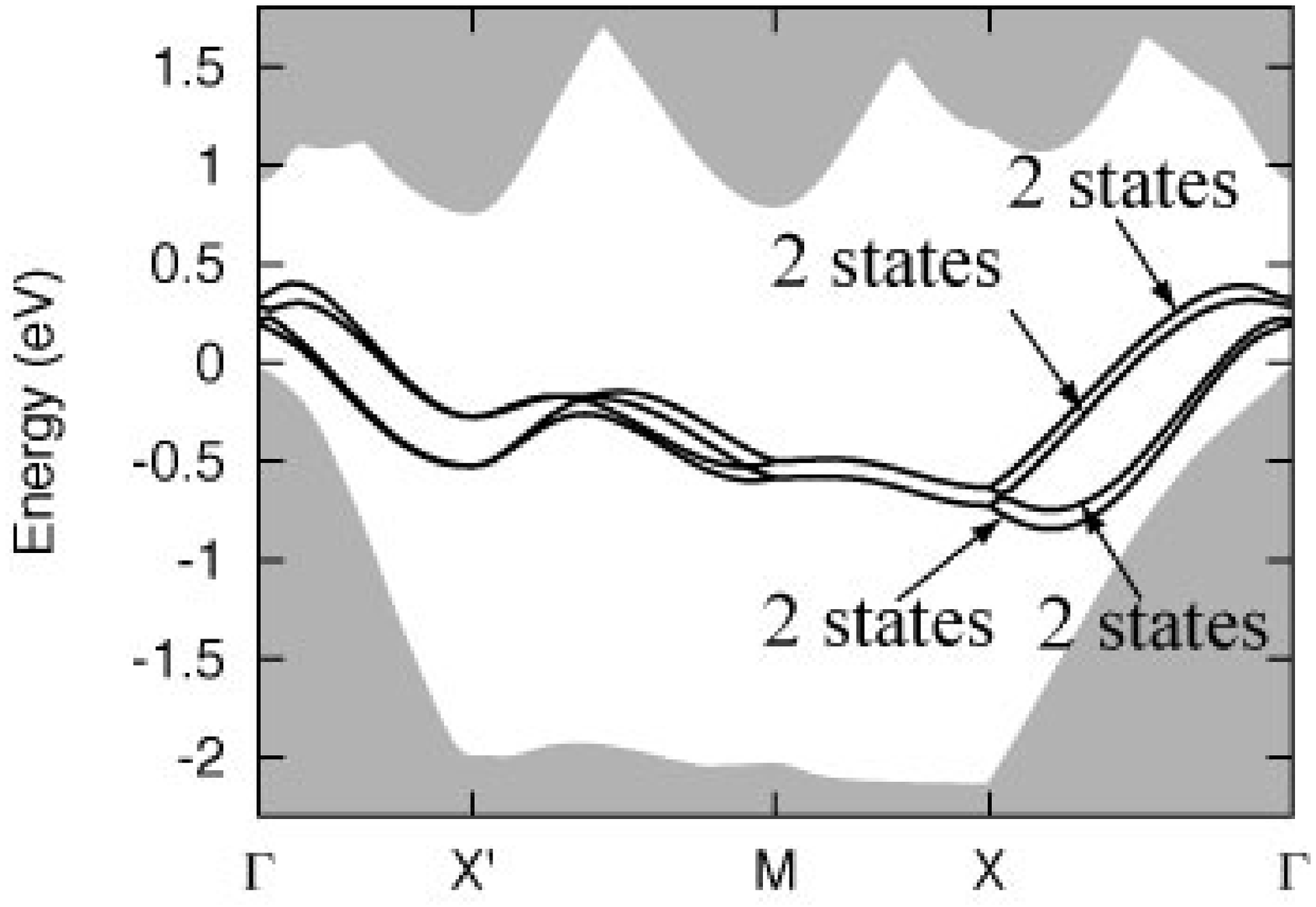}\\
  (b) Ge (100)\\
  \includegraphics[width=\figsize]{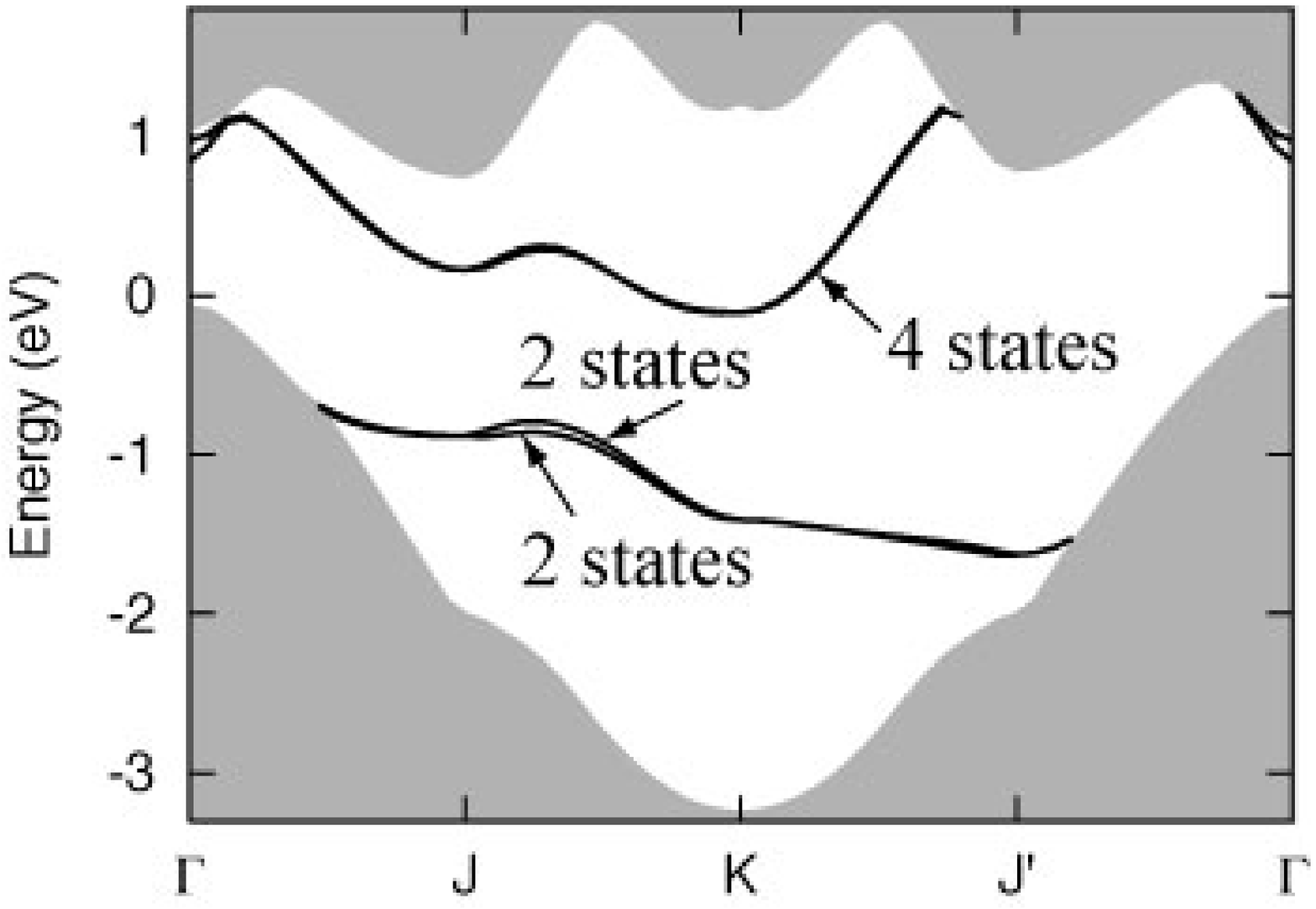}\\
  \caption{Energy spectrums of Ge that have ideal surface (110) and (100).
  The notations of points in relative surface Brillouin zone are
  according to Ivanov {\it et. al.}\cite{Ivanov_1980}.
  The number in the figures is the degeneracy of the edge states are
  eight including spin degrees of freedom.
  At (100) surface, edge states disappear for special values of $\mbf{k}_{\|}$.}
  \label{band_ge}
   \end{center}
 \end{figure}
 \begin{figure}[hbtp]
   \begin{center}
  (a) Si (111)\\
  \includegraphics[width=\figsize]{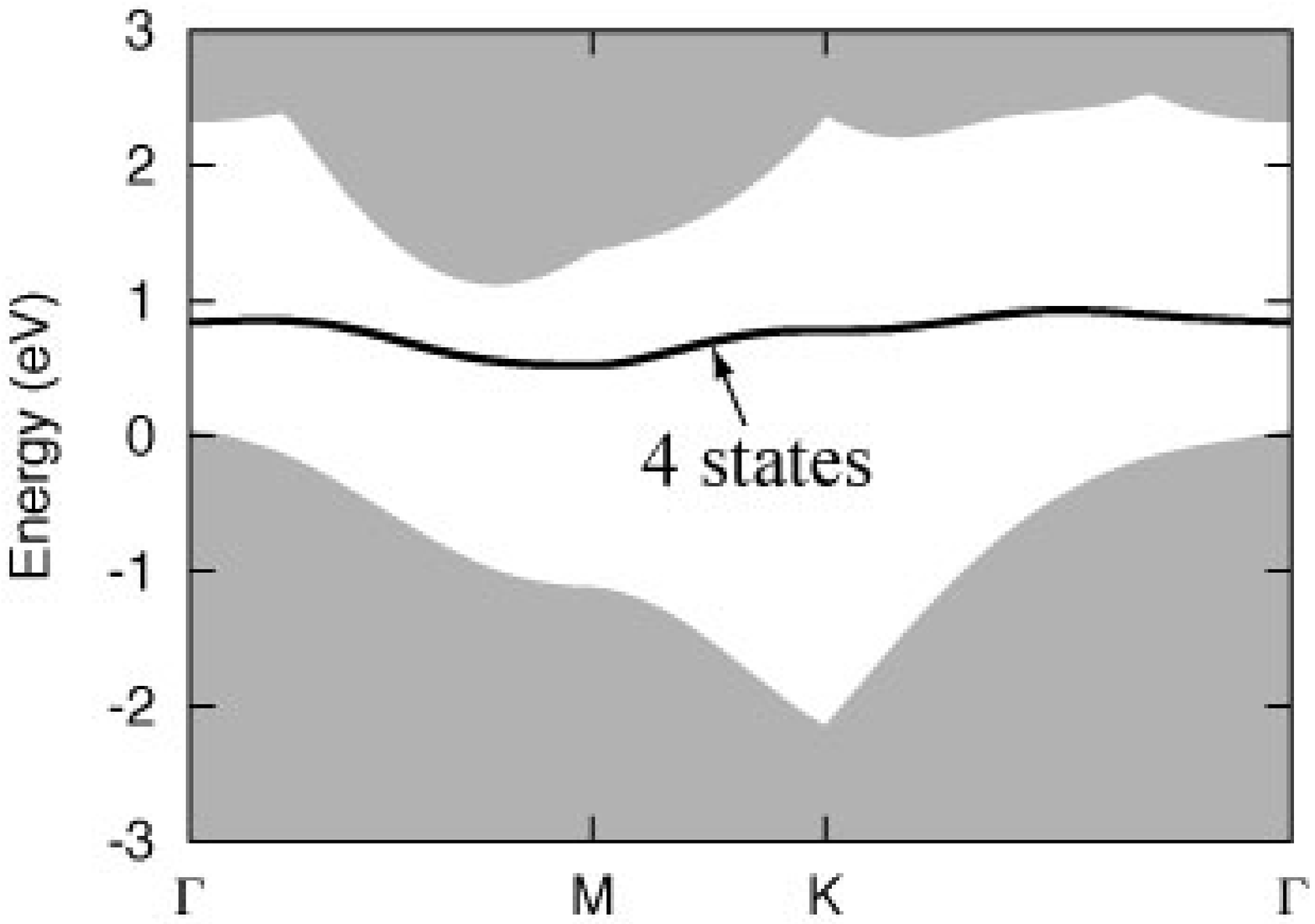}\\
  (b) Si (110)\\
  \includegraphics[width=\figsize]{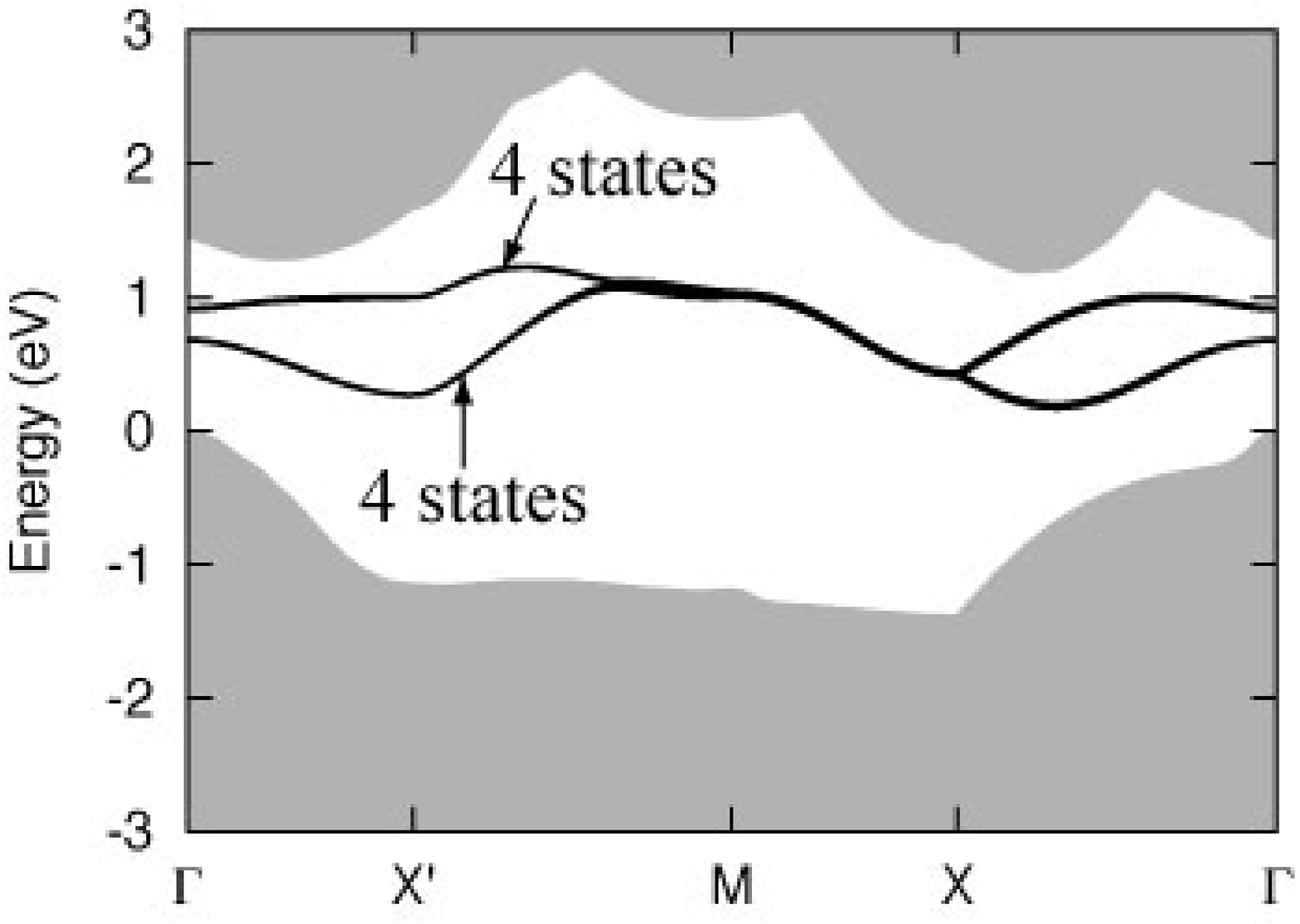}\\
  (c) Si (100)\\
  \includegraphics[width=\figsize]{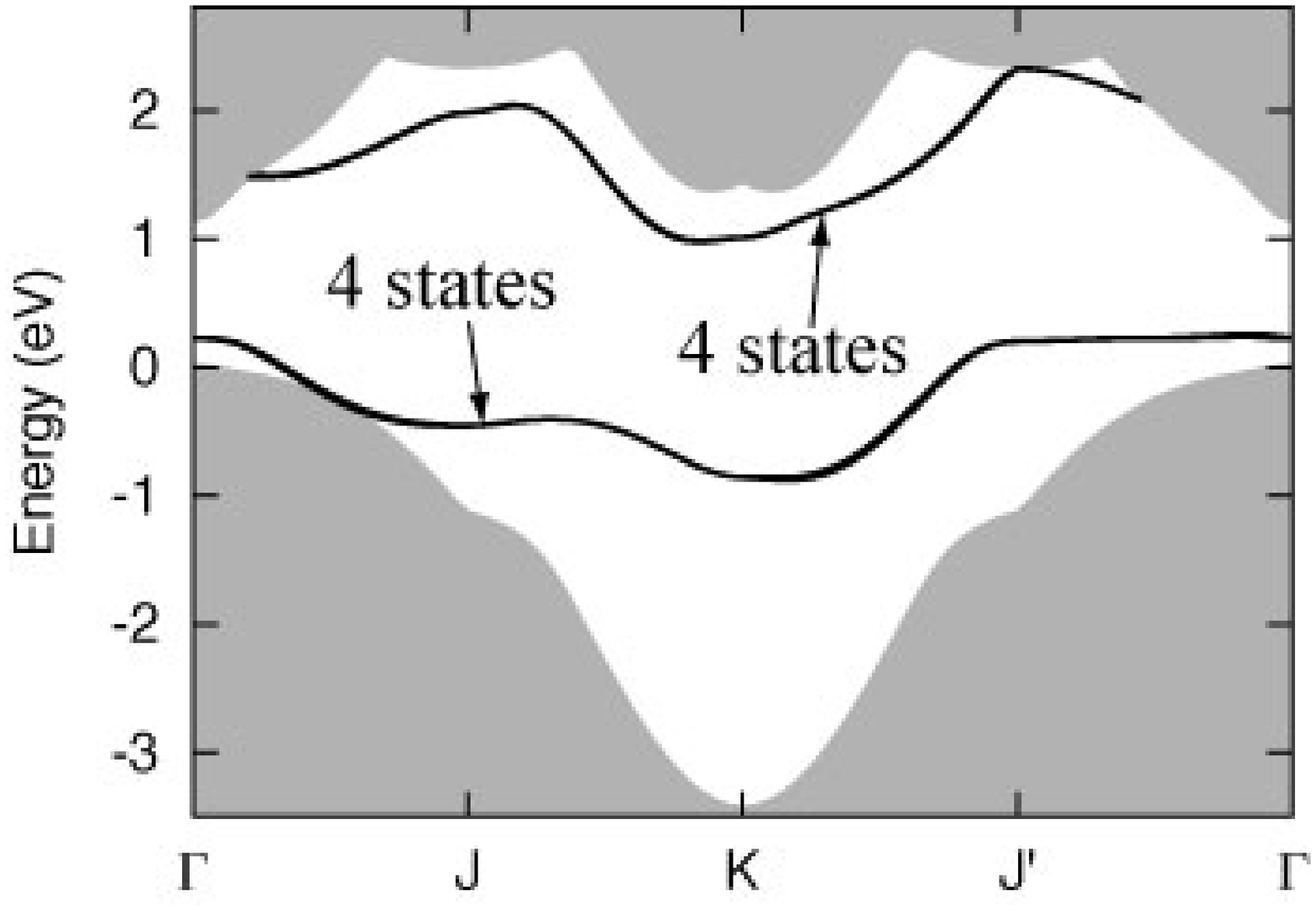}\\
  \caption{Energy spectrums of Si that have ideal surface (111), (110),
  and (100).
  At (111) surface and (110) surface, four and eight surface states
  appear for each $\mbf{k}_{\|}$ including spin degrees of freedom, relatively.
  At (100) surface, eight surface states appear including spin degrees
  of freedom for special values of $\mbf{k}_{\|}$.
  }
  \label{band}
\end{center}
\end{figure}

 Here, we summarize results of the other surfaces of Si and Ge
 studied in the same manner as Ge (111).
 As shown in Figs.~\ref{band_ge} and \ref{band}
 in addition to in Fig.~\ref{fig.1} (b),
 each energy band diagram of edge states of Ge (110), Ge (100), Si (111), Si (110), and Si(100)
 shows different energy dependence.
 These edge states are all doubly degenerated due to two sides of surface.
 In Ge (110), Si (110), Ge (100), and Si (100),
 value of energy split between two kind of edge states
 is roughly estimated as bonding states and anti-bonding states
 of two dangling bonds on the surface.
 Especially, the edge states merged into the bulk bands for (100) cases
 and we cannot define the number of edge states $n_e$.

 The Berry phase $\gamma$ turns out to be quantized by the anti-unitary symmetry
 and succeed in clarification of the three surfaces for Si, Ge, and the decoupled covalent-bond model
 after the adiabatic transformation while $n_e$ fails for the (100) surface,
 as summarized in Table \ref{table_edge}.
 Since it is topological quantity,
 the result does not depend on the detail of the tight binding
 parameters unless the gap remains open.
 Then, Ge and Si with the same surface are equivalent
 from a viewpoint of the Berry phases.

 After the adiabatic deformation described above,
 the winding number $\mathcal{W}$ for the chiral symmetric Hamiltonian
 classifies the different surfaces.
 The results are also summarized in Table \ref{table_edge}.
 The sign of $\mathcal{W}$ determines whether the loop $\mathcal{L}$ is clockwise or anti-clockwise,
 and becomes important to classify $(110)$ and $(001)$ surfaces.  
 The latter is identical to $(100)$ surface.
 We note that $n_e$ can be defined for the chiral symmetric Hamiltonian
 of Ge (100) and Si (100),
 because edge states emerges from the bulk bands into the gap in the adiabatic transformation.
 However,
 we cannot allow such adiabatic transformation if we consider $n_e$ as a topological invariant.

 The decoupled covalent-bond model clarifies the meaning of topological invariants
 because edge states are identified as dangling bonds.
 Then, $n_e$ of each materials refers the number of dangling bonds,
 and a non-trivial ($\pi$) value of $\gamma$ indicates
 the link where a dangling bond exists.

 \begin{table}[htbp]
  \begin{tabular}{|c|c|c|c|}\hline
   Surface & $\gamma=\pi$ & $n_e$ per spin & $\mathcal{W}_{\alpha}\neq0$ \\\hline
   Si, Ge, CB (111) & $\gamma_{1}$ & $n_e=2$ & $\mathcal{W}_{1}=1$
   \\
   Si, Ge, CB (110) & $\gamma_{1}, \gamma_{\bar{4}}$ & $n_e=4$ & $\mathcal{W}_{1}=1,\mathcal{W}_{4}=-1$
   \\
   Si, Ge (100) & $\gamma_{1}, \gamma_{2}$ & --- & $\mathcal{W}_{1}=1,\mathcal{W}_{2}=1$
   \\
   CB (100) & $\gamma_{1}, \gamma_{2}$ & $n_e=4$ & $\mathcal{W}_{1}=1,\mathcal{W}_{2}=1$ \\ \hline
  \end{tabular}
  \caption{We have summarized material, their surface,
  the Berry phase $\gamma$ that have nontrivial value $\pi$,
  number of edge states $n_{e}$ per spin,
  and winding numbers $\mathcal{W}_{\alpha}$ that are non-zero.
  Here, we dropped a spin index because the results are spin independent:
  $\gamma_{\alpha}=\gamma_{\alpha\uparrow}=\gamma_{\alpha\downarrow}$
  and $\mathcal{W}_{\alpha}=\mathcal{W}_{\alpha\uparrow}=\mathcal{W}_{\alpha\downarrow}$.
  CB denotes the decoupled covalent-bond model defined in text.}
  \label{table_edge}
 \end{table}

 \paragraph{Conclusion}
 In conclusion,
 we have studied topological invariants
 at some ideal surfaces of Ge, Si
 described by the tight binding model with
 Slater and Koster's parameter.
 Following the previous study on the quantized Berry phase\cite{Hatsugai_2006},
 we have defined the surface-specified Berry phase.
 It should be emphasized that the symmetry required for quantization of
 the Berry phase is the inversion and conjugate complex symmetry while
 it is the particle-hole symmetry and conjugate complex symmetry in the
 previous study.
 It will suggest a possibility for studies of the other materials
 without the particle-hole symmetry.

 From the adiabatic transformation,
 Ge, Si, and the decoupled covalent-bond model with the same surface are
 topologically equivalent
 from a viewpoint of the Berry phase.
 In other words, these topological invariants
 successfully deduced the simple picture of dangling bonds
 from the complicated tight binding model with a spin-orbit coupling.
 Moreover,
 the results show that the Berry phase is successful for all surfaces of Ge, Si
 and provide information about a position of dangling bonds.

 Finally, we comment on GaAs.
 For the (111) surface,
 there are four edge states $n_e=4$.
 The value of energy split of edge bands is
 roughly estimated as difference between
 on site potential of two atoms.
 The site potential makes it difficult the adiabatic transformation to
 the simple model as is the case with Si, Ge (100).
 The Berry phase cannot be quantized
 because GaAs has no inversion symmetry.
 However,
 the Berry phase defined by another twist angle can overcome the difficulty
 if a anti-unitary symmetry for it's quantization is founded in GaAs,

  \acknowledgments
 We acknowledge discussions with T. Hirano.
 This work was supported by Grant-in-Aid from
 the Ministry of Education, No. 17540347 from JSSP, No.18043007 on
 Priority Areas from MEXT and the Sumitomo Foundation.
 Some of numerical
 calculations were carried out on Altix3700BX2 at YITP in Kyoto
 University.
 
\end{document}